\documentclass[a4paper, 10pt, twocolumn]{article}

\usepackage{layout}
\usepackage{amsmath}
\usepackage{mathtools}
\usepackage{multirow}
\usepackage{caption}
\usepackage{graphicx} 
\usepackage{subcaption}
\usepackage{algorithm}
\usepackage{color}
\usepackage{algpseudocode}
\usepackage{hyperref}
\usepackage{float} 
\hypersetup{
  colorlinks,
  citecolor=black,
  linkcolor=black,
  urlcolor=black}

\usepackage[numbers,sort&compress]{natbib}
\usepackage[a4paper, margin=15mm]{geometry}

\usepackage{setspace}
\usepackage{longtable}

\usepackage[symbol]{footmisc}

\usepackage{booktabs}
\usepackage{multirow}
\usepackage{graphicx}

\usepackage{tabularx}   
\usepackage{makecell}   
\usepackage{array}

\newcolumntype{L}{>{\raggedright\arraybackslash}X}
\newcolumntype{C}{>{\centering\arraybackslash}X}
\newcolumntype{R}{>{\raggedleft\arraybackslash}X}

\usepackage[affil-it]{authblk}
\title{Cornering in the Water: An Investigation of Dolphin Swimming Performance}

\begin{document}



\author[1]{Mingkai Xia\thanks{The two authors have equal contributions in this paper}}
\author[1]{Junhan Zhang\protect\footnotemark[1]}
\author[1]{Ningshan Wang}
\author[1]{Gabriel Antoniak}
\author[2]{Nicole West}
\author[1]{Ding Zhang}
\author[1]{Kenneth Alex Shorter\thanks{Corresponding author: Associate Professor, Mechanical Engineering, University of Michigan, 2350 Hayward, Ann Arbor, MI 48109, \texttt{kshorter@umich.edu}}}

\affil[1]{Department of Mechanical Engineering, University of Michigan, Ann Arbor, MI 48109, USA}
\affil[2]{Dolphin Quest Oahu, Honolulu, HI 96816, USA}

\date{}

\maketitle

\begin{abstract}

Marine mammal biomechanics research has focused on straight-line swimming at consistent speeds, resulting in a lack of knowledge about how animals select movement strategies to balance cost vs performance during tasks like cornering. In this work we examine performance, maneuverability and cost tradeoffs for bottlenose dolphins (\textit{Tursiops truncatus}) during prescribed swimming. During the task, animals completed two straight-line sections of swimming with a cornering event (180-degree turn).  Movement kinematics were measured with a biologging tag (speed, orientation, and depth), and used to estimate the path of the animal during cornering events using a dead reckoning approach. A hydrodynamic model was used to estimate thrust power and energetic cost during lap swimming. Three animals performed the same swimming task, but the path, cornering strategy, and speed varied between individuals. From the kinematic analysis, TT02 was the fastest lap swimmer, with the highest average lap speed, along with the largest energetic cost. TT01 selected a strategy that reduced energetic cost by sacrificing task performance; the animal took about 1.5 times longer to finish each lap compared to TT02 (36 s vs 23 s). TT03 swam more slowly than TT02 (28 s vs 23 s), but at a 50\% reduction in cost per lap. The improved efficiency seen in TT03’s movement strategy was the result of reducing transient costs during the lap. This included selecting cornering trajectories that balanced trade-offs between distance traveled and speed loss during the turn. Results from this work provide new insight into maneuverability and movement strategies that the dolphins adopt to balance performance and cost during movement, and provide new knowledge for the design and control of bio-inspired marine robotic systems.


\end{abstract}

\section{Introduction}\label{sec:Intro}
Performance, energetic cost, and environmental or behavioral context are competing objectives that must be balanced when selecting movement strategies for both biological and engineered systems. New knowledge about how biological systems weigh these costs, and how movement strategies may vary between individual animals, will inform the design and operation of engineered systems. Dolphins are excellent swimmers, with high propulsive efficiencies over a range of swimming speeds \citep{fish2014measurement}. In contrast, conventional single-screw propellers used to drive engineered systems in the marine environment have good propulsive efficiencies (as high as 0.85) but only in a narrow speed band about a design point \citep{fish2006limits}. To this end, there has been interest in the development of bio-inspired propulsion systems that emulate the foil-like fluke motion and swimming patterns of dolphins.


The design, control, and evaluation of robotic systems benefit from measurements and estimates of dolphin swimming biomechanics. For example, dolphins have been observed to select swimming gaits and speeds that reduce energetic cost \citep{zhang2023dynamics},~\citep{gabaldon2022tag}. These observed movement profiles have informed the control objectives for robotic systems. Li et al. have developed a robust motion control system informed by measured dolphin kinematics to select depth, speed, and heading of a dolphin-inspired robot to achieve various tasks in the field \citep{li2024robust}. Additionally, Wang et al. present simulation results for a three-dimensional (3-D) path planning method for a robotic dolphin during gliding descents \citep{wang20193}. Researchers have also developed bio-inspired energy-saving strategies for high-performance underwater vehicles by incorporating a dynamic model to reduce the cost of transport (Li et al. \citep{li2024towards}). Measurement and analysis of movement profiles from dolphins provide knowledge that has been used to improve the design and performance of the bio-inspired engineered systems, but the marine environment makes investigating dolphin swimming biomechanics challenging. 

Camera-based sensing strategies used to measure kinematics are limited when the dolphins are underwater, and hydrodynamic forces acting on the animals can not be measured directly. As such, work tends to use a combination of kinematic measurements and model-based force estimates to investigate swimming biomechanics. Movement kinematics have been measured using overhead and underwater cameras \citep{gabaldon2022computer} and wearable sensors \citep{kaidarova2023wearable}. Using camera data, Fish and Rohr \citep{fish1999review} found that the movement patterns dolphins select during fluking return the body to a stable pose during bouts of movement. Camera data have also been used to characterize body pose during swimming, and to estimate propulsive forces during low amplitude swimming \cite{zhang2020simulated}. This approach tends to capture only a few continuous fluke strokes when the animals are within the camera field of view, highlighting the challenge of measuring a dolphin's swimming path in a three-dimensional marine environment.  Camera-based localization approaches provide reliable positions for dolphins at the surface, but environmental conditions (animal depth, water quality, glare) can limit animal visibility, resulting in sparse planar positional estimates. In contrast, biologging tags (wearable sensors for animals) provide continuous high-fidelity kinematic measurements (depth, orientation, speed, acceleration) from the animals. These kinematic measurements can be used to construct dead-reckoning tracks \citep{wensveen2015path}. Sensor drift affects the accuracy of the estimate over time, but drift error can be greatly reduced by limiting the duration and step-size of the tracks \citep{sharp2014sensor}. 

Estimates of propulsive forces and power are also important for the investigation of dolphin swimming performance. Both direct and indirect models have been used to estimate propulsive forces during swimming. Direct approaches have focused on modeling the fluke as a foil-like propulsor with measured kinematics used as inputs to move the foil at biologically relevant speeds and orientations with respect to the direction of motion ~\citep{xargay2023low},~\citep{yu2011control},~\citep{antoniak2023estimating}. Indirect approaches use hydrodynamic models of drag acting on the body with measured speed and acceleration to estimate thrust force generated during swimming \citep{gabaldon2022tag}. Both of these approaches use measured kinematics as model inputs to estimate hydrodynamic forces during swimming. The measured speed can then be used with the estimated propulsive forces to estimate movement energetics (work, power, and cost of transport)  during swimming \citep{zhang2023dynamics},~\citep{gabaldon2022tag},~\citep{fish1993power}.

Studies investigating movement energetics have tended to focus on animal performance during consistent speed straight-line swimming. For example, Antoniak et al. \citep{antoniak2023estimating} used a model-based approach to investigate propulsive efficiency and found that animal-specific parameters, like fluke area, can significantly impact swimming mechanics. That study focused on estimates of planar swimming at consistent speeds and depths. However, animal movement often involves rapid changes in direction in response to the environment, and dolphins may select different movement strategies depending on the task. For example, animals may minimize movement costs in response to increased path variability \citep{wilson2021path}.  Quantitative analyses of turning maneuverability in freshwater turtles demonstrated that smaller and younger turtles exhibited higher angular velocities and tighter turning radii \citep{stevens2018ontogeny}. To date, the investigation of maneuverability and performance of marine animals during tasks like turning remains limited. Recent work has analyzed four distinct turning modes for robotic dolphins and demonstrated that coordinated head–flipper maneuvers achieve the minimum turning radius and maximum angular velocity, providing design and control guidelines for robotic dolphins \citep{yang2022research}. An improved understanding of how dolphins prioritize cost vs performance during tasks that require a high degree of maneuverability is important for engineers developing biologically inspired underwater robots.

In this article, we combine localization with a tag-based biomechanics analysis to investigate movement patterns during a prescribed swimming task with a 180-degree turn. Biologging tags recorded movement kinematics, including the depth, orientation, and speed through the water. We leveraged the measured kinematics to estimate the acceleration (normal and tangential) and the swimming path of the dolphins using the dead-reckoning method. Subsequently, we applied a hydrodynamic modeling approach (See Gabaldon et al. \citep{gabaldon2022tag}) to estimate propulsive power, work, and cost of transport (COT).  Finally, we analyzed the estimated swimming biomechanics and energetics to investigate how the individual animals selected paths and movement strategies to balance energetic cost and performance during the task.

\section{Experiment Setup}\label{sec:Setup}

\begin{figure}[htbp]
    \centering
    \includegraphics[width=0.49\textwidth]{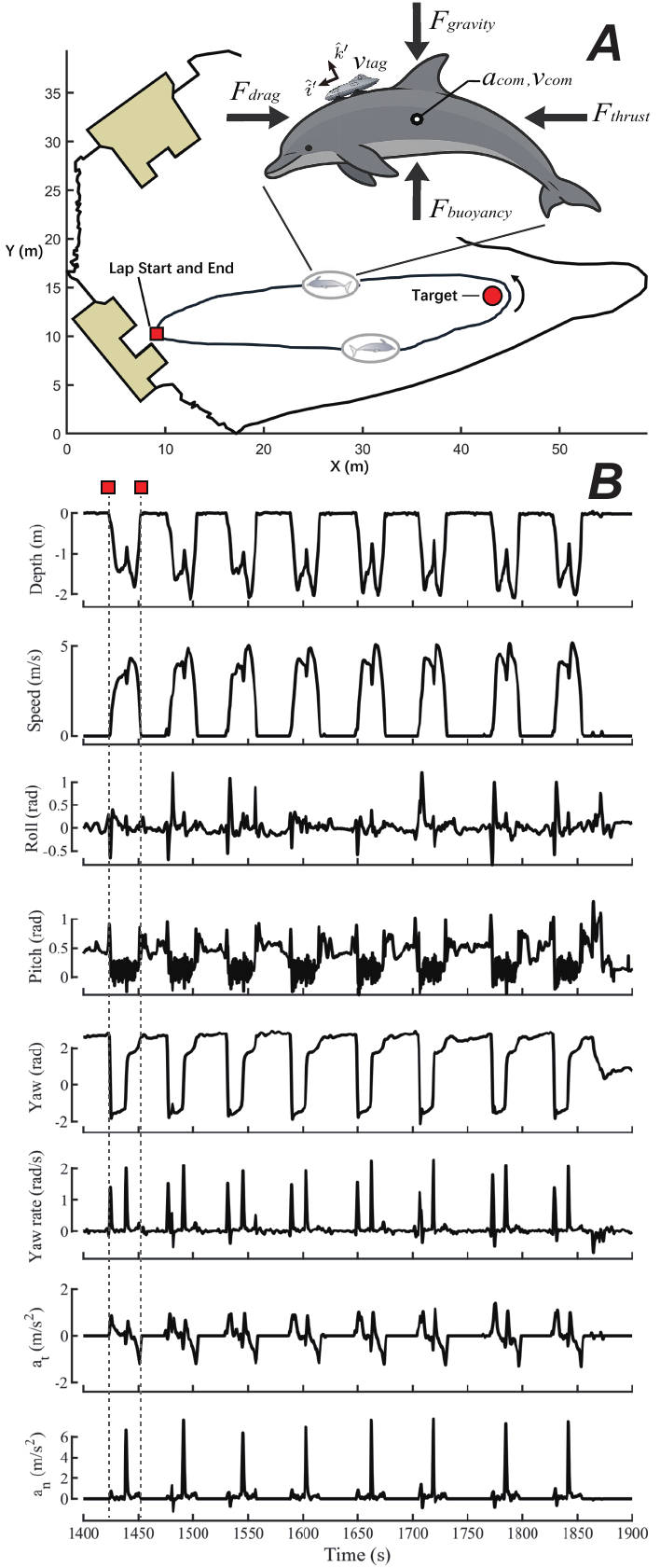}
    \caption{\textbf{A}: Free-body diagram of a swimming dolphin that illustrates the body-fixed reference frame of the tag/animal system and the speed sensor, along with an example lap track from an experimental trial. Trials consisted of 7 to 10 laps, with the dolphin starting from the station and then swimming underwater around a target at the far end of the lagoon before returning to the starting position. \textbf{B}: Measured and calculated kinematic parameters for the eight laps completed during the representative trial (TT03). The start and end of an individual lap have been identified using the red squares and grey dashed lines.}
    \label{fig: Lap trial}
\end{figure}

Experimental trials were conducted with bottlenose dolphins (\textit{Tursiops truncatus}) in a shallow lagoon environment at  Dolphin Quest Oahu, HI. During the experiment, dolphins performed a prescribed swimming task that consisted of repeated laps to a target placed at the far side of the lagoon shown in Fig.~\ref{fig: Lap trial}.  During the first half of the lap, the dolphins start from rest and accelerate to a consistent speed before a cornering event where the animals have to make a 180-degree change in direction. Following the turn, the animals accelerate to a consistent speed before gliding to a station at the end of the lap. The animals station for 5 -10 seconds (approximately 3 respiration events) before repeating the lap up to ten times. During the trials, movement kinematics (depth, speed, acceleration, angular velocity, and heading) were measured using a biologging tag (MTag) \citep{lauderdale2021bottlenose}. The tags were secured to the animals using four silicone suction cups, and the dolphins are trained to wear the MTags prior to the study. Marine mammal specialists used positive behavioral reinforcement to train the animals to complete the swimming task while wearing the tags. Data were collected from three animals that ranged in size from 143 to 245 kg and length 2.2 to 2.5 m. All experimental work was approved by the Institutional Animal Welfare and Use Committee at the University of Michigan.

\section{Modeling and Analysis}\label{sec:Method}

\subsection{Movement Kinematics}
Tag measurements were used to calculate parameters for the localization (orientation, normal, and tangential components of body acceleration) and kinetics (thrust power) estimates.  MTag sensors include a 3-axis accelerometer, 3-axis gyroscope, and  3-axis magnetometer \citep{lauderdale2021bottlenose}. Angular velocity, acceleration, and heading were recorded at a sampling rate of 50 Hz and filtered with the Madgwick filter \citep{madgwick2010efficient}, to estimate orientation (pitch, roll, and heading). Speed through water was measured using a magnetic micro-turbine mounted outside the tag housing and a Hall effect sensor. Environmental temperature and pressure were also recorded. Depth and forward speed were recorded at a sampling rate of 5 Hz and were filtered as presented in \citep{gabaldon2021contextualized} shown in Fig.~\ref{fig: Lap trial}. Vector parameters are presented in the world reference frame.

\subsubsection{Acceleration of the Body}
\emph{Forward} (tangential) acceleration \(a_{t}\) was calculated from the smoothed speed measurement using a finite‐difference approach:
\begin{equation}
a_{t,i}
\approx
\frac{v_{i+1}-v_{i-1}}{2\,\Delta t}\
\end{equation}
where \(a_{t,i}\) is the tangential acceleration at sample \(i\) ($i$ is a positive integer), \(v_{i}\) is the linear speed in the body-fixed frame at sample \(i\) smoothed using the moving-average method with a window size of $1 s$, \(v_{i+1}\) and \(v_{i-1}\) are the speeds at the next and previous time‐steps, respectively. \(\Delta t\) is the constant time interval between samples.

Because the length of the lap (30 m) was much greater than the depth of the animal during the task (less than 2 m), swimming trajectories were assumed to be planar for the analysis. Planar angular speed, \(\omega_i\), of the animals during the laps was defined in the world \(x\)--\(y\) plane and calculated using a finite‐difference approach with the smoothed yaw‐angle time series:
\begin{equation}
\omega_{i}
\approx
\frac{\psi_{i+1}-\psi_{i-1}}{2\,\Delta t}\,
\end{equation}
where \(\omega_{i}\) is the angular speed at sample \(i\), \(\psi_{i}\) is the yaw angle at sample \(i\) smoothed using the moving-average method with a window size of $1 s$, \(\psi_{i+1}\) and \(\psi_{i-1}\) are the yaw angles at the next and previous time‐steps, respectively.

Finally, the \emph{normal} (centripetal) acceleration was calculated using the linear and angular speeds: 
\begin{equation}
a_{n,i}
=
\omega_{i}\,v_{i}\,
\end{equation}
where \(a_{n,i}\) is the normal (centripetal) acceleration at sample \(i\). 

\subsection{Tag-Based Modeling of the Swimming Biomechanics}

The modeling approach presented in this work is based on \citep{gabaldon2022tag}. It is assumed that the dynamics of the animal can be modeled as a rigid body and that dolphins can modulate their buoyancy to compensate for gravity. The free-body diagram is shown in Fig.~\ref{fig: Lap trial}. The hydrodynamic forces acting on the body were simplified into net thrust and drag forces acting along the line of motion of the body. The positive direction for both thrust and drag force is defined along the dolphin’s swimming direction. Their relation can be represented by Newton's 2nd law:
\begin{equation}
    m^{\prime} a_{\mathrm{t}}=F_{\text {thrust }}+F_{\text {drag }}
\end{equation}
where \(a_{\mathrm{t}}\) was assumed to be the acceleration of the center-of-mass (COM) and \(m^{\prime}\) is the total effective mass (mass of the animal and the added mass of the water displaced by the body) defined below:
\begin{equation}
m^{\prime}=m+m_{\mathrm{add}}
\end{equation}
where m is the mass of the animal and \(m_{add}\) is the added mass of fluid assumed to be \(m_{add}\) = 0.4 $\rho V$ at the fluid density $\rho$ = 1030 kg/m\(^3\) and dolphin volume $V$. The drag force acting on the dolphin \(F_{\text {drag}}\) was estimated using a depth-dependent drag model \citep{gabaldon2022tag}: 
\begin{equation}
F_{\text {drag }}=-0.5 \rho A_{\mathrm{S}} C_{\mathrm{D}} \gamma v_{\mathrm{COM}}^2 
\end{equation}
where \(A_s\) = 0.08\(m^{0.65}\) denotes the surface area, \(C_D\) = 16.99\(Re^{-0.47}\) denotes the normalized drag coefficient \citep{fish1993power}, $\gamma$ represents the depth-dependent coefficient \citep{hertel1966structure} accounting for wave drag when the animal swims near the surface, and \(v_{COM}\) is the speed of the center-of-mass through the water. The thrust power is defined by:
\begin{equation}
P_{\text {thrust }}=F_{\text {thrust }} v_{\mathrm{com}}=\underbrace{m^{\prime} a_{\mathrm{t}} v_{\mathrm{com}}}_{\text {Inertial }}-\underbrace{F_{\text {drag }} v_{\mathrm{com}}}_{\text {Drag }} 
\end{equation}
where \(F_{\text {thrust}}\) is the thrust force generated by the dolphin. By integrating the formulas mentioned above, we can derive the equation for thrust power:
\begin{equation}
P_{\text {thrust }}=\underbrace{(m+0.4 \rho V) a_{\mathrm{t}} v_{\text {com }}}_{\text {Inertial }}+\underbrace{0.5 \rho A_{\mathrm{s}} C_{\mathrm{D}} \gamma v_{\text {com }}^3}_{\text {Drag }} 
\end{equation}
The following formulas are used to present the predicted relationship between thrust power and linear velocity:
\begin{equation}
\widehat{P}_{\text {thrust }}\left(v_{\mathrm{com}}\right)=a_1 v_{\mathrm{com}}^{a_2}
\end{equation}
\begin{equation}
\widehat{P}_{\mathrm{t}, \mathrm{nd}}\left(v_{\mathrm{com}}\right)=b_1\left(v_{\mathrm{com}} / L\right)^{b_2}
\end{equation}
where \(a_1\), \(b_1\), \(a_2\), \(b_2\) are scalars calculated by non-linear curve fitting process and \(L\) is the dolphin's body length. These scalars are shown to be positive, resulting in a positive relationship for both thrust power and non-dimensional thrust power with linear velocity. The energetic cost during the dolphin's movement is defined as follows:
\begin{equation}
\mathrm{COT}=\frac{P_{\text {thrust }} /\left(\eta_{\mathrm{ms}} \eta_{\mathrm{sp}}\right)+P_{\mathrm{RMR}}}{m v_{\mathrm{com}}} 
\end{equation}
where \(P_{RMR}\) represents the metabolic power of the dolphin at rest, \(\eta_{ms}\) = 0.25 represents the efficiency of converting chemical energy to mechanical energy in mammals \citep{massaad2007up}, accounting for energy loss during this process. \(\eta_{sp}\) = 0.85 indicates the efficiency of converting internal power into external propulsive power allowing the movement in water. A predicted model of the cost is defined as follows:
\begin{equation}
 \widehat{\mathrm{COT}}=\frac{\widehat{P}_{\text {thrust }}\left(v_{\mathrm{com}}\right) /\left(\eta_{\mathrm{ms}} \eta_{\mathrm{sp}}\left(v_{\mathrm{com}}\right)\right)+P_{\mathrm{RMR}}}{m v_{\mathrm{com}}},
\end{equation}
where \(\widehat{\mathrm{COT}}\) is the predicted energetic cost.

Dolphin speed was normalized to body length by dividing the speed of the dolphin by the body length for comparison between dolphins with different properties (length):
\begin{equation}
    v_{\mathrm{normal}} = v_{\mathrm{com}}/L
\end{equation}
where \(v_{\mathrm{normal}}\) is the speed normalized to body length (body-length speed). Thrust power was also normalized to a non-dimensional form \(P_{t,nd}\) to make comparisons between dolphins with different properties (length and mass):
\begin{equation}
P_{\mathrm{t}, \mathrm{nd}}=P_{\text {thrust }} /\left(m g^{1.5} L^{0.5}\right),
\end{equation}
where g is the gravitational acceleration constant (9.81 m/\(s^2\)). The normalization constant, $mg^{1.5}L^{0.5}$, for dolphin TT01 is 7174.2\,$W$, TT02 is 11983.1\,$W$, and TT03 is 6492.5\,$W$ (Table.~\ref{tab:performance_metrics}).

The thrust work was also normalized by dividing the thrust work by the normalization constant $mg^{1.5}L^{0.5}$ to make comparisons between dolphins with different properties (length and mass):
\begin{equation}
W_{\mathrm{t}, \mathrm{normal}}=W_{\text {thrust }} /\left(m g^{1.5} L^{0.5}\right)
\end{equation}
where $W_{\text {thrust }}$ denotes the thrust work calculated by applying \(W_{\text{thrust}} \approx \sum_{i=1}^{N} P_{\text{thrust},i} \,\Delta t\). Here, $P_{\text{thrust},i}$ denotes the thrust power of the dolphin at sample $i$, and \(\Delta t\) denotes the sample time step.
\\
\subsubsection{Localization Estimation}
The path used by the animal during the swimming task was estimated using a dead-reckoning approach \citep{wensveen2015path}. This method is a navigation technique employed to ascertain the position of a moving object by estimating its direction and distance traveled from a known starting point. Position updates are based on the previous location estimate along with the current heading and speed in the direction of movement. The heading vector and the speed in the direction of motion were calculated from measured kinematics. Because the movement of the animals was assumed to be planar the animal's velocity was projected into the \(x-y\) plane using the dolphin's pitch angle.  





{The lagoon boundary was defined to establish a reference for dolphin movement in the environment. The lagoon boundary was created using Google Earth and imported into MATLAB using the Mapping Toolbox to convert coordinates from latitude and longitude into metric distances from an origin. The following equation was used to update the position estimate (\(x,y\)) of the dolphin within the lagoon,
\begin{equation}
    \Vec{p}_{t+1}=\Vec{p}_{t}+v\widehat{n}\Delta t
\end{equation}
where \(\widehat{n}\) is the unit heading vector in \(x-y\) plane, \(\Delta t\) the time between position estimates (reciprocal of the sampling frequency), \(\Vec{p}_{t}\) is the current position vector at time t (t is an integer), and \(\Vec{p}_{t+1}\) is the position vector at time \(t+1\).  The following equation was used to project the three-dimensional velocity into the two-dimensional x-y plane: 

\begin{equation}
    v = v_\textup{meas}\cos(\theta)
\end{equation}
where \(v\) is the speed of the dolphin in 
\(x-y\) plane, \(\theta\) denotes the dolphin's pitch angle and \(v_{meas}\) represents the scalar speed in the body-fixed coordinate system measured by the MTag. \\
The unit heading vector was calculated as follows:
\begin{equation}
    \widehat{n} = \cos(\psi)\hat{i} + \sin(\psi)\hat{j}
\end{equation}
where \(\psi\) is the dolphin's yaw angle, \(\hat{i}\) is the unit vector in \(x\) direction and \(\hat{j}\) is the unit vector in \(y\) direction.
}




\subsubsection{Cornering Analysis}
The instantaneous radius of curvature during cornering events was calculated from the dead-reckoning trajectory using the following method \citep{stewart2013essential}:


\begin{equation}
R_i 
= 
\frac{\bigl(\dot x_i^2 + \dot y_i^2\bigr)^{3/2}}
     {\bigl|\dot x_i\,\ddot y_i - \dot y_i\,\ddot x_i\bigr|}
\end{equation}
where $R_i$ is the instantaneous radius of curvature at time‐step $i$ ($i$ is a positive integer),  $\dot x_i \approx \dfrac{x_{i+1}-x_{i-1}}{2\Delta t}$,  \quad
$\dot y_i \approx \dfrac{y_{i+1}-y_{i-1}}{2\Delta t}$, $\ddot x_i \approx \dfrac{x_{i+1}-2x_i+x_{i-1}}{(\Delta t)^2}$, \quad
$\ddot y_i \approx \dfrac{y_{i+1}-2y_i+y_{i-1}}{(\Delta t)^2}$, $\Delta t$ is the constant time interval between successive samples, $x_i$ and $y_i$ are coordinates of the trajectory in the world frame at the $i$th time sample.  



\subsection{Event Detection and Lap Normalization}

Active fluking periods were defined using the measured pitching motion of the animal. Active swimming was further divided into periods of transient and consistent speed swimming using the calculated acceleration of the animal shown in Fig.~\ref{fig:Event_detection}. To compare movement kinematics (position, speed, acceleration) and energetics (mechanical work and power) of the individual animals, data were normalized to lap percentage using the cornering event as a common reference point. Kinematic data were used to identify the lap start (\(t_{\mathrm{s}}\)), lap end (\(t_{\mathrm{e}}\) ), and the cornering event (\(t_{\mathrm{c}}\)), Fig.~\ref{fig:Event_detection}. At the lap start (\(t_{\mathrm{s}}\)), the speed, forward acceleration and thrust power begin to increase rapidly, and the pitch angle starts to oscillate. At the lap end (\(t_{\mathrm{e}}\) ),  speed, forward acceleration, and thrust power return to zero, Fig.~\ref{fig:Event_detection}. The time of the cornering event (\(t_{\mathrm{c}}\)) was identified using the maximum of the normal acceleration and defined the mid-point of the normalized lap ( \(50\%\) ). The start and end of the turn were defined when the normal acceleration dropped to 55\% of its maximum value before and after the cornering event. Data were normalized to percent lap using the following approach:

\begin{equation}
t_\textup{\%lap}=
\begin{cases}
\dfrac{t_\textup{end}}{2t_\textup{c}}\,\cdot t & \text{if } t \le t_\textup{c} \\[10pt]
t_\textup{end} - \dfrac{ t_\textup{end}}{2(t_\textup{end} - t_\textup{c})}\,(t_\textup{end} - t) & \text{if } t > t_\textup{c}
\end{cases}
\end{equation}
where \(t_{\mathrm{\%lap}}\) is the percentage-lap time after the normalization process, \(t_{\mathrm{end}}\) is the ending time of the lap before normalization, \(t_{\mathrm{c}}\) is the time of the cornering event, which is also the time before normalization when the normal acceleration reaches maximum, and \(t\) is the time recorded by the MTag. All the times are in seconds. Then the \text{Percentage Lap} is calculated using the following equation:

\begin{equation}
\text{Percentage Lap}=\frac{t_\textup{\%lap}}{\textup{max}(t_\textup{\%lap})} 
\end{equation}
where \(\textup{max}(t_\textup{\%lap})\) denotes the maximum percentage-lap time for a lap, which corresponds to the lap's end time after normalization.


\begingroup
\setlength{\textfloatsep}{6pt}

\begin{figure}[t!]
    \centering
    \includegraphics[width=0.49\textwidth]{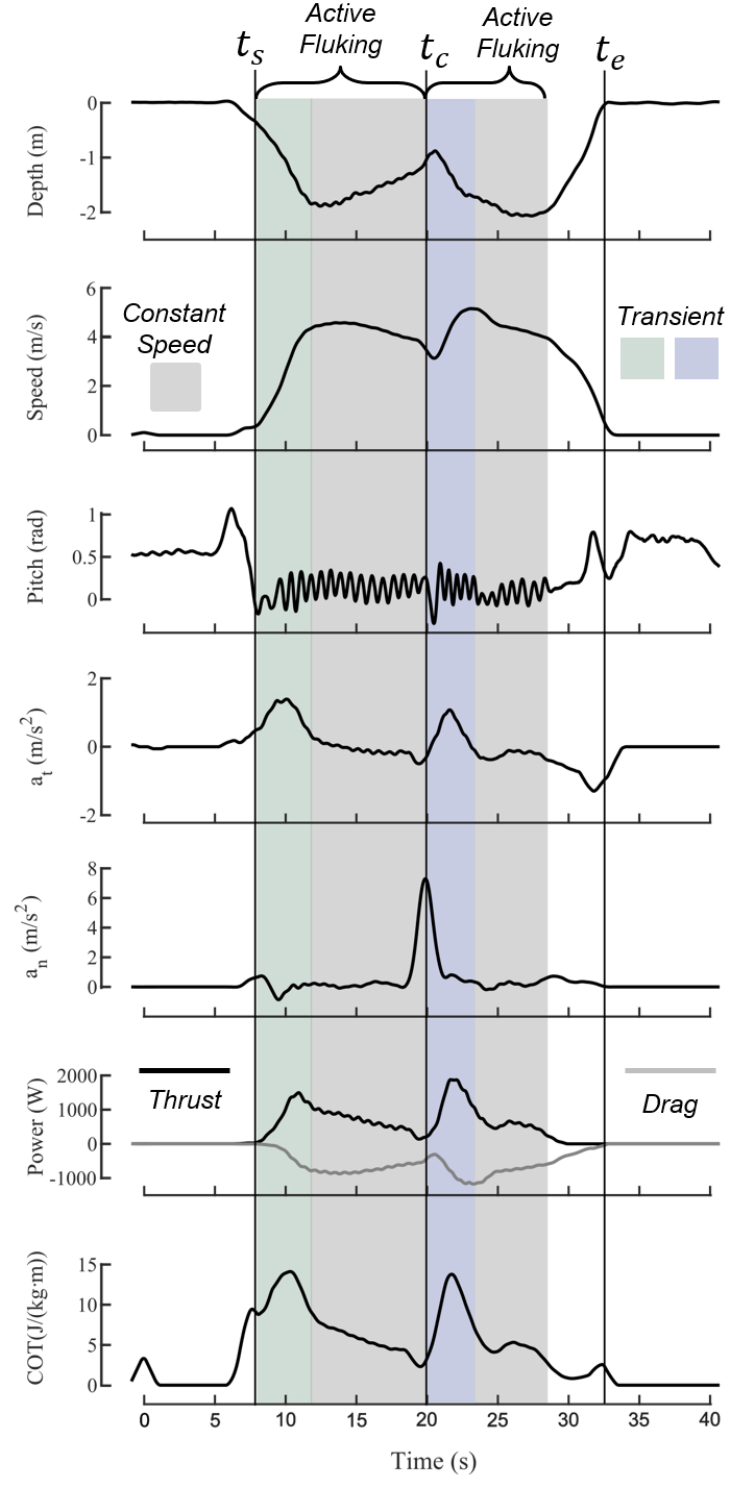}
    \caption{Measured and estimated kinematic data for dolphin TT03 from a representative lap. Features in the data were used to identify: 1) lap start (\(t_{s}\)), 2) the cornering event (\(t_{c}\)), and 3) lap end (\(t_{e}\)). Active fluking during the lap was identified using the speed and pitching measurements. The green and blue shading denotes transient periods of swimming, and the grey shading denotes the consistent speed swimming.}
    \label{fig:Event_detection}
\end{figure}

\endgroup

\section{Results}\label{sec:Results}
Data were collected from 30 trials  (TT01 - 12 trials, TT02 - 8 trials, and TT03 -  10 trials) with a total of  221 laps used for the analysis. Representative movement kinematics (depth, speed, orientation, and acceleration) from a lap trial are presented in Fig.~\ref{fig: Lap trial}. During the 8-minute trial, the dolphin completed eight laps and reached peak speeds around 4 m/s during the return leg of the lap. Fig.~\ref{fig:Event_detection} presents a representative lap, where the animal starts from rest and accelerates to a speed around 4 m/s in about 5 seconds (transient swimming). The animal maintains this speed up to the cornering event, around 9 seconds. At the turn, the animal's speed decreases to around 3 m/s and the normal acceleration peaks at around 6 m/$\textup{s}^2$. Coming out of the turn, the animal accelerates from 3 to 5 m/s and then swims for around 5 seconds before the animal glides to station. Average lap duration ranged from 23.1 s for TT02 to 35.7 s for TT01. Periods of transient swimming during both the outgoing and return portions of the laps were between 2.1 s (TT01) to 3.2 s (TT02) for all animals.  Periods of active fluking during the return lap ranged from 8.5 s (TT02) to 11.9 s (TT01), less than half of what was observed during the outgoing lap. Periods of gliding were observed for all the animals during the return and ranged from 3.1 s (TT02) to 8.6 s (TT01). (Table. \ref{table: event duration}).

\begin{table*}[t]
  \centering
  \scriptsize
  \setlength{\tabcolsep}{3pt}
  \renewcommand{\arraystretch}{1.15}
  \caption{Average event durations for the transient (Trans.), consistent-speed (CS), and active-fluking (AF) phases, together with cornering timing for the three dolphins for all laps. The location of the start and end of the average cornering event are presented in Fig.~\ref{fig:dr_xyt}}
  \label{tab:merged_one}
  \resizebox{\textwidth}{!}{%
  \begin{tabular}{lcccccccccccc}
    \toprule
    \multirow{2}{*}{Dolphin} &
    \multirow{2}{*}{Lap Duration~(s)} &
    \multicolumn{3}{c}{Outgoing — Duration~(s)} &
    \multicolumn{4}{c}{Return — Duration~(s)} &
    \multicolumn{3}{c}{Cornering — Timepoints~(s)} &
    \multirow{2}{*}{Cornering - Duration~(s)} \\
    \cmidrule(lr){3-5}\cmidrule(lr){6-9}\cmidrule(lr){10-12}
    & &
    Trans. & CS & AF &
    Trans. & CS & AF & Gliding &
    Start & $t_c$ & End \\
    \midrule
    TT01 & $35.7\pm5.1$ &
    $2.1\pm1.1\,(6\%)$ & $13.1\pm5.0\,(37\%)$ & $15.2\pm2.5\,(43\%)$ &
    $2.4\pm2.0\,(7\%)$ & $9.5\pm3.2\,(26\%)$ & $11.9\pm2.8\,(33\%)$ & $8.6\pm2.9\,(24\%)$ &
    $14.4\pm2.5$ & $15.2\pm2.5$ & $16.0\pm2.5$ &
    $1.6\pm0.2$ \\
    TT02 & $23.1\pm2.6$ &
    $3.2\pm1.1\,(14\%)$ & $8.3\pm3.5\,(36\%)$ & $11.5\pm1.3\,(50\%)$ &
    $2.7\pm0.8\,(12\%)$ & $5.8\pm2.4\,(25\%)$ & $8.5\pm2.2\,(37\%)$ & $3.1\pm0.5\,(13\%)$ &
    $10.7\pm1.3$ & $11.5\pm1.3$ & $12.3\pm1.2$ &
    $1.6\pm0.2$ \\
    TT03 & $28.3\pm3.4$ &
    $2.9\pm1.1\,(10\%)$ & $11.4\pm3.8\,(40\%)$ & $14.3\pm1.7\,(50\%)$ &
    $2.1\pm1.5\,(8\%)$ & $8.2\pm2.3\,(29\%)$ & $10.3\pm2.0\,(37\%)$ & $3.7\pm0.8\,(13\%)$ &
    $13.6\pm1.7$ & $14.3\pm1.7$ & $15.0\pm1.7$ &
    $1.4\pm0.1$ \\
    \bottomrule
  \end{tabular}}
  \label{table: event duration}
\end{table*}

\begin{table*}[!t]
\centering
\begingroup
\setlength{\tabcolsep}{0.5pt}      
\renewcommand{\arraystretch}{1.08} 
\fontsize{7}{8.5}\selectfont       
\renewcommand\theadfont{\fontsize{7}{8.5}\selectfont} 
\caption{Dolphin parameters and performance metrics (lap averages) for the three dolphins during all trials.}
\label{tab:performance_metrics}
\begin{tabularx}{\textwidth}{C C C C C C C C C C C C C}
\toprule
\thead{Dolphin} &
\thead{Mass\\(kg)} &
\thead{Body\\Length\\ (m)} &
\thead{Norm.\\Const.\\(W)} &
\thead{$P_{\mathrm{RMR}}$\\(W)} & 
\thead{Total\\Travel\\Distance\\(m)} &
\thead{Peak\\Speed\\(m/s)} &
\thead{Mean\\Speed\\(m/s)} &
\thead{Peak\\Power\\(kW)} &
\thead{Mean\\Power\\(kW)} &
\thead{Peak\\Angular\\Velocity\\(rad/s)} &
\thead{Average\\Cornering\\Radius\\(m)} &
\thead{Total\\Thrust\\Work\\(kJ)} \\
\midrule
TT01 & 156.2 & 2.24 & 7174.2 & 347.9 &
$82 \pm 6$ & $2.9 \pm 0.4$ & $2.0 \pm 0.2$ & $0.4 \pm 0.1$ & $0.17 \pm 0.06$ &
$1.3 \pm 0.1$ & $1.3 \pm 0.3$ & $0.15 \pm 0.04$ \\
TT02 & 244.7 & 2.54 & 11983.1 & 442.9 &
$78 \pm 8$ & $5.2 \pm 0.7$ & $3.3 \pm 0.4$ & $2.3 \pm 0.7$ & $1.0 \pm 0.3$ &
$1.4 \pm 0.1$ & $1.9 \pm 0.8$ & $1.4 \pm 0.7$ \\
TT03 & 142.6 & 2.20 & 6492.5 & 317.6 &
$78 \pm 4$ & $4.2 \pm 0.6$ & $2.8 \pm 0.3$ & $1.1 \pm 0.3$ & $0.4 \pm 0.1$ &
$1.8 \pm 0.1$ & $1.0 \pm 0.2$ & $0.5 \pm 0.1$ \\
\bottomrule
\end{tabularx}
\endgroup
\end{table*}


\subsection{Path Selection and Cornering}
\begin{figure*}[htbp]
    \centering
    \includegraphics[width=\textwidth]{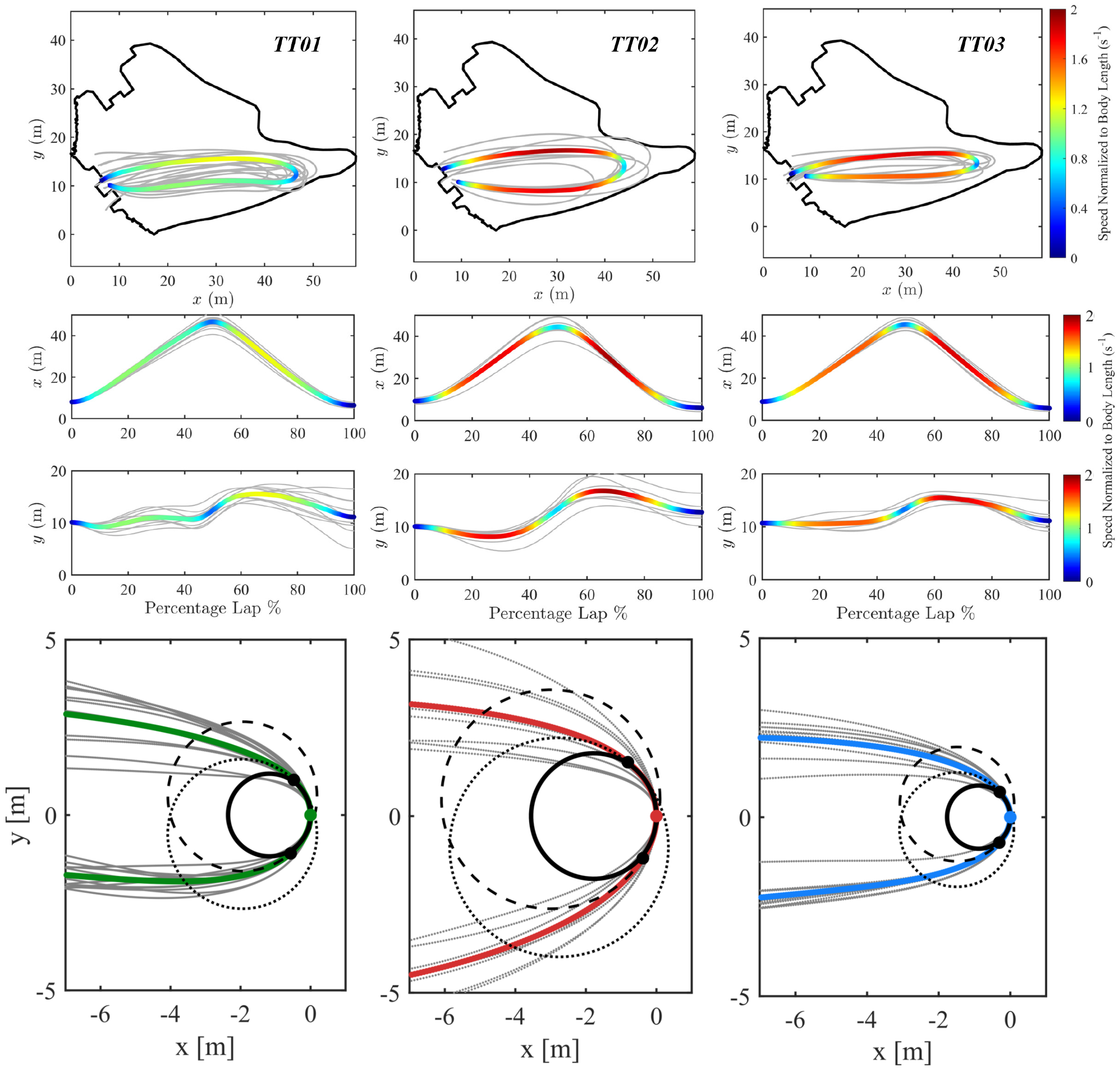}
    \caption{\textbf{Top:} Estimated path and measured speed during the lap swimming trials shown with the boundary of the main lagoon. Individual (grey lines) and trial average paths are presented along with the average lap speed (heat map). \textbf{Middle:} - Horizontal (x) and vertical (y) position of the animal shown as a percentage of the lap with the cornering event occurring at the midway point (50 percent). Speed was normalized to body length. \textbf{Bottom:} - Swimming paths for each of the animals aligned at the cornering event. Data from each trial is presented with the average path for a trial presented in grey and the trial average presented as a bold, colored line. The circle of best fit for the trajectory is presented 30 percent before (dashed), at (solid line), and 30 percent after (dotted) the cornering event.}
    \label{fig:dr_xyt}
\end{figure*}

Animal speed and orientation were used to estimate the position of the animal during the lap. Average results from each trial, along with the trial average, were used to investigate differences in movement kinematics observed during the laps. Each trial average represents approximately eight laps. TT02 had the fastest average lap speed (\(3.3 \pm 0.4\) m/s), while TT01 had the slowest average speed (\(2.0 \pm 0.2\) m/s) shown in Fig.~\ref{fig:dr_xyt}. All three animals swam fastest during the straight line sections of the lap, but TT01 swam slower than the other two animals for the trial presented in Fig.~\ref{fig:dr_xyt}. The path selected by the animals varied as well, with the average distance traveled ranging from \(78 \pm 8\) m for TT02, \(78 \pm 4\) m for TT03, to \(82 \pm 6\) m for TT01. The animals all swam approximately the same horizontal ($x$) distance (38 m to 40 m), but TT02 selected a path that spanned a greater vertical distance (TT01: 6 m, TT02: 9 m, and TT03: 5 m). 

\begin{figure*}[htbp]
    \centering
    \includegraphics[width=0.9\textwidth]{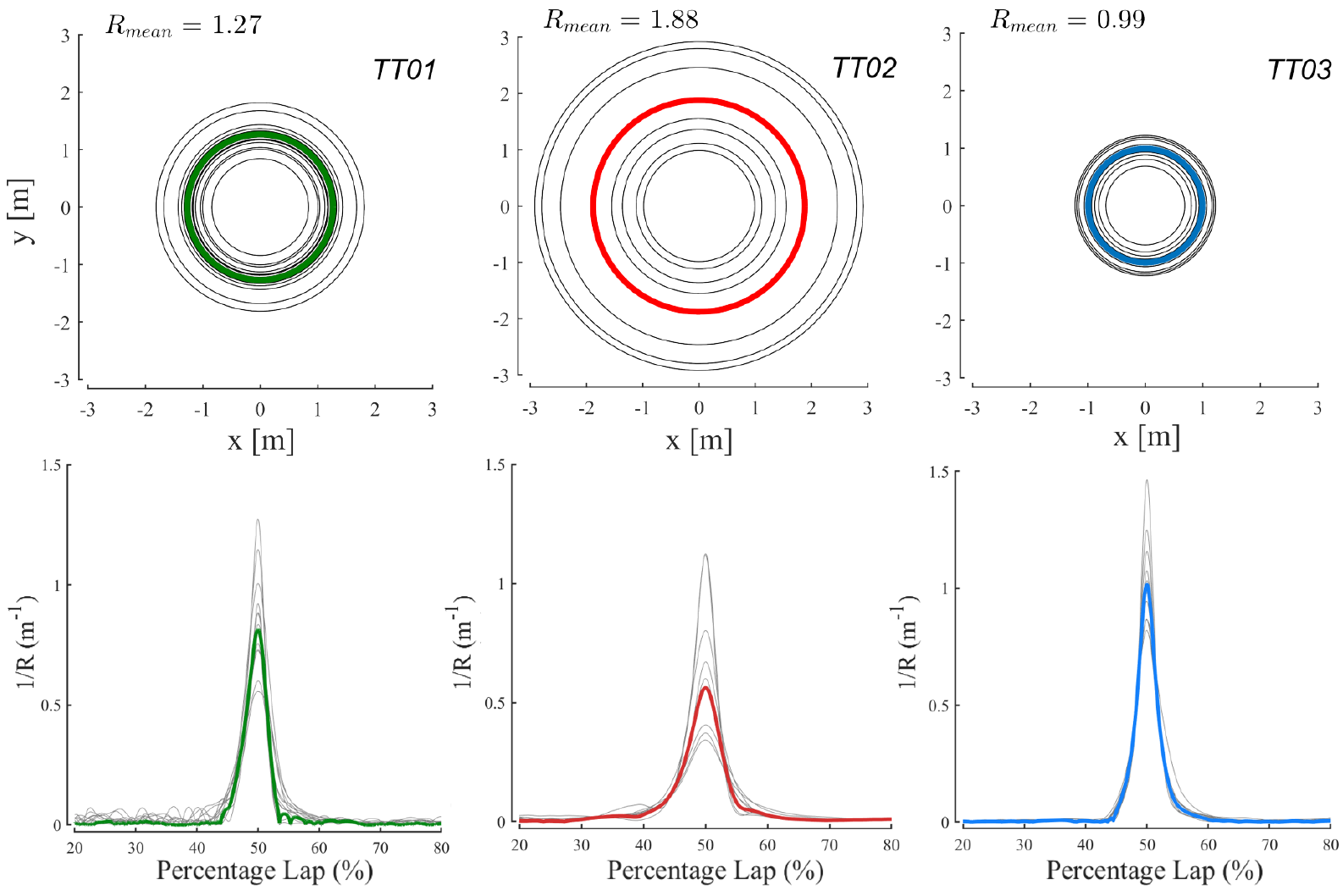}
    \caption{\textbf{Top:} Trial average radius at the cornering event for each of the animals. \textbf{Bottom:} How the radius of curvature changes along the swimming path as a percentage of the lap with 50$\%$ corresponding to the cornering event.}
    \label{fig:R_circle}
\end{figure*}

\begin{figure*}[htbp]
    \centering
    \includegraphics[width=0.85\textwidth]{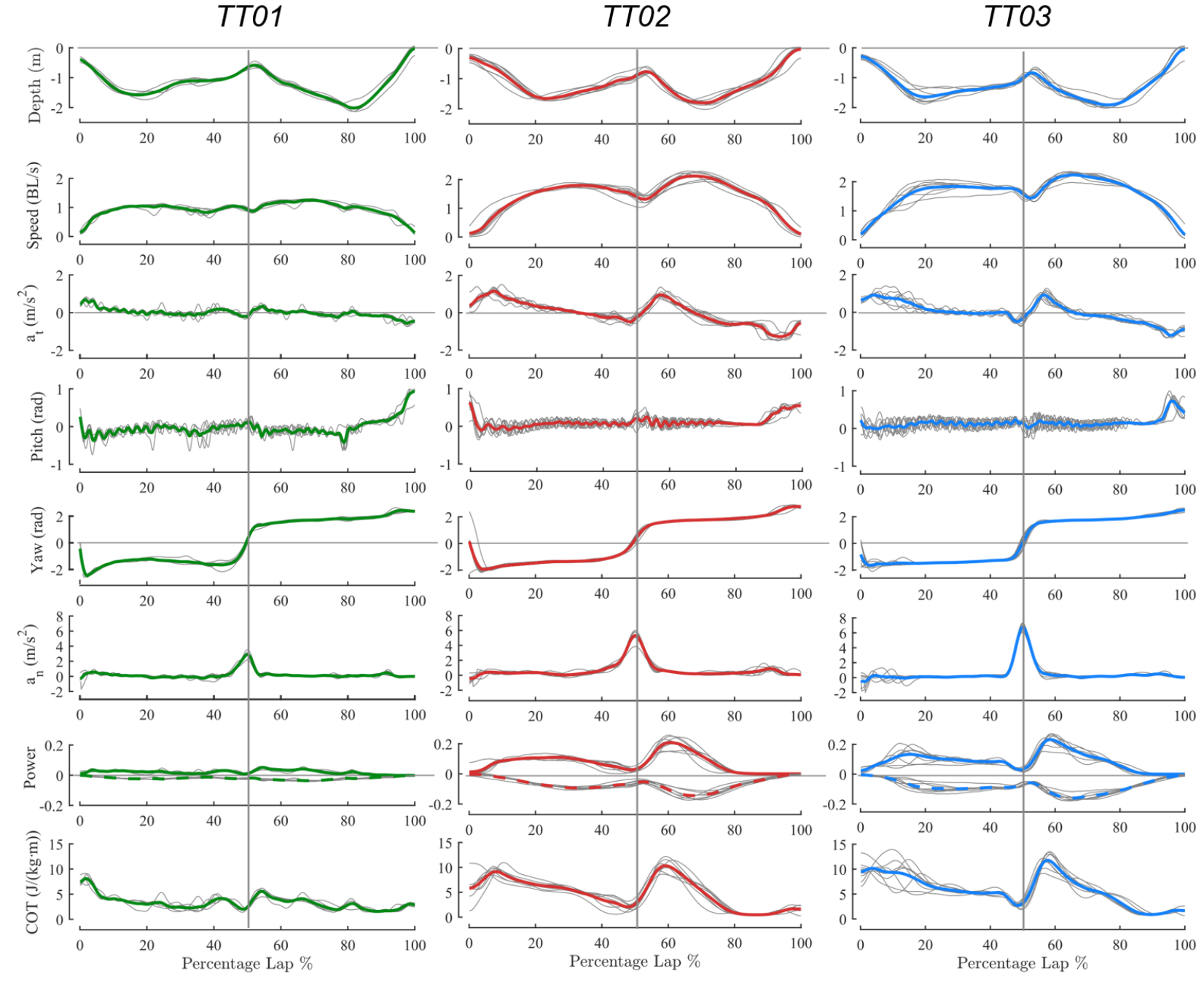}
    \caption{Key kinematic (depth, speed, acceleration) and kinetic (thrust and drag power) parameters from the animals for one trial (\~{8} laps). Data are normalized to percentage lap, and the cornering event occurs at 50\% of the lap. The individual (grey lines) and average (colored bolded line) data for the results are presented in each subfigure.}
    \label{fig:avg_laps}
\end{figure*}

To investigate cornering behavior, the estimated path of the animal was used to compute the instantaneous radius of curvature for each trial, Fig.~\ref{fig:dr_xyt}. The estimated position of the animals at the cornering event was used to align position estimates to compare paths selected by the animals during cornering,  Fig.~\ref{fig:R_circle}. In addition to the longest path, TT02 also had the largest average radius of curvature (1.88 m) at the cornering event compared to TT01 (1.27 m) or TT03 (0.99 m), Fig.~\ref{fig:R_circle}.  TT02 had the most variability on a trial-to-trial basis, with the average radius of curvature varying from 1.0 m to 2.9 m. The larger trial average radius for curvature also indicates that TT02 had the lowest peak normal acceleration at the cornering event 0.5 m/$\textup{s}^{2}$ compared to  TT01 (0.75 m/$\textup{s}^{2}$) or TT03 (1.0 m/$\textup{s}^{2}$), Fig~\ref{fig:avg_laps}.




\begingroup
\setlength{\textfloatsep}{6pt}  

\begin{figure}[t!]
    \centering
      \includegraphics[width=0.48\textwidth]{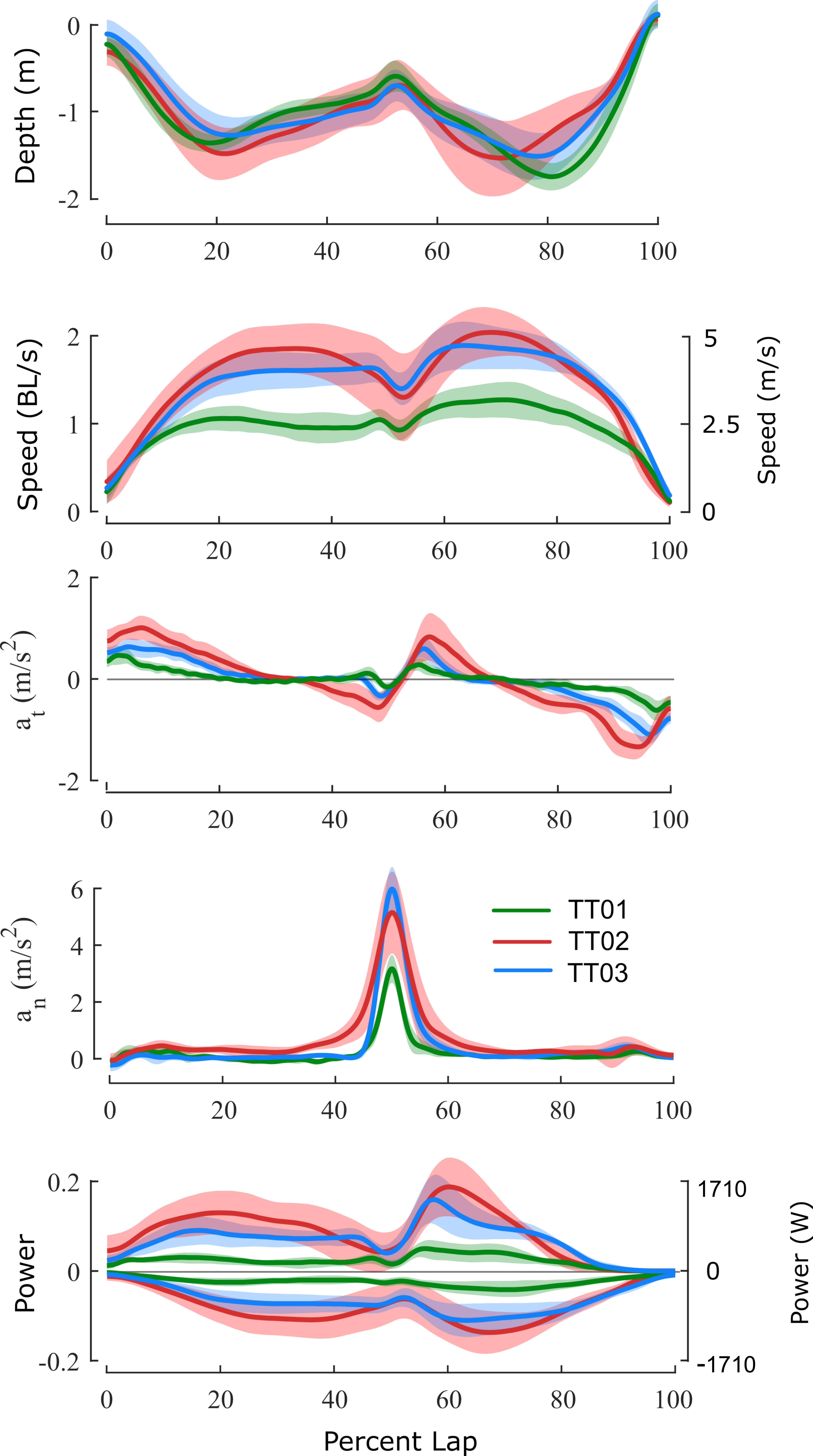}
    \caption{Comparison of average kinematic (depth, speed and acceleration) and kinetic parameters
    (drag and thrust power) for all trials. Thrust power is positive and drag power is negative for the three animals. The dimensional values on the right-hand axis are calculated using the average body length and the mean normalization constant.}
    \label{fig:avg_trails}
\end{figure}

\endgroup

\subsection{Energetic Cost Analysis}
Measured kinematics (depth, speed, acceleration, and orientation) and estimated kinetics (thrust force, drag force, and propulsive power) were normalized and then averaged on a per-trial basis for the three animals, Fig.~\ref{fig:avg_laps}.  Trends in the kinematic data were broadly comparable for the dolphins. The depth during the trials is deepest during the return leg of the lap, and shallowest just after the cornering, even for all three animals. The average pitching during the trials indicates that all three animals tended to fluke continuously during the outgoing leg of the lap and through the cornering event before an extended period of gliding during the last 20 percent of the lap. The heading data indicates major changes in direction during the lap. At the start of the animals reorient towards the far end of the lagoon as they begin the lap. The dolphins then maintain a consistent heading until they approach the turn at the far side of the lagoon (around 45 percent of the trial). After changing direction after the cornering event, the animals again maintain a consistent heading during the return leg of the lap. Periods of positive acceleration in the direction of travel are largest at the start of the lap and after the cornering event. Changes in normal acceleration peak at the cornering, start around 5 percent before the event, and continue for around 5 percent after the event. 

TT01 swam the slowest (around 2.0 m/s) and accelerated gradually (around 0.3 m/$\textup{s}^{2}$) to a consistent swimming speed during the laps, which resulted in smaller power estimates throughout the trial compared to the other two dolphins, Fig~\ref{fig:avg_trails}. The largest average power output estimated for TT02 (2.3 kW) and TT03 (1.1 kW) occurs around 60 percent of the lap when the animals are accelerating to the fastest speeds observed during the trials.  TT02 had the fastest average swimming speed, reaching 5.2 m/s during the return leg of the lap, as well as larger peak linear accelerations during both legs of the lap (around 1 m/$\textup{s}^{2}$). This resulted in a larger power output throughout the trial than either TT01 or TT03. For all animals, power estimates dropped after peaking as the animals stopped fluking and glided to a stop at the station, Fig.~\ref{fig:avg_trails}.  Cost of transport estimates for TT01 are lower than TT02 or TT03, again reflecting the slower swimming speeds selected by TT01. Cost of transport estimates were highest when the animals were accelerating to the consistent swimming speed at 10 and 60 percent of the lap, Fig.~\ref{fig:avg_laps}.

\begin{figure}[t!]
    \centering
    \includegraphics[width=0.49\textwidth]{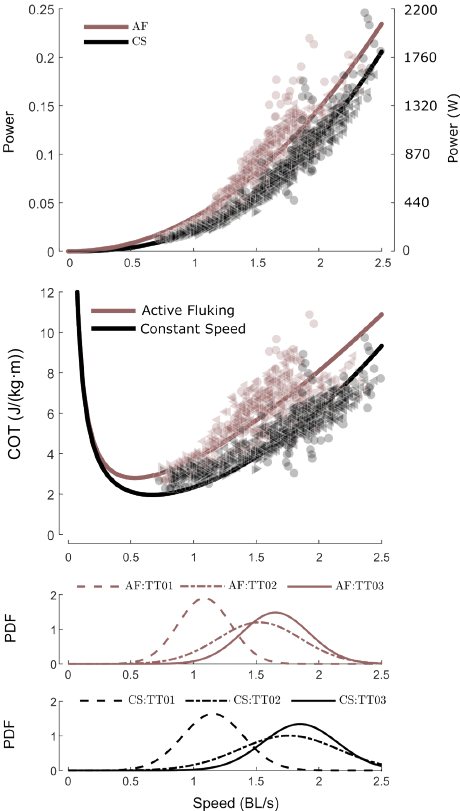}
    \caption{\textbf{Top:} Cost of transport for all active fluking during the lap swimming (AF) and the shorter sections of constant speed swimming (CS). The transient movements during the trials results in the increased cost of transport. \textbf{Middle:} Power for the active fluking and constant speed portions of the data during the trial. The additional power required to accelerate the animal to the constant swimming speed is reflected in the higher power requirements over the range of observed swimming speeds. \textbf{Bottom:} Speed distributions.}
    \label{fig:AF VS CS COT}
\end{figure}

Average power and speed were calculated for the periods of active fluking and consistent speed swimming for the outgoing and return portions of the laps. These data were used to compare thrust power and cost of transport for the three animals over a range of swimming speeds, Fig.~\ref{fig:AF VS CS COT}. Functional fits to the average thrust power estimates during both periods of active fluking (\(\widehat{P}_{t,\mathrm{nd}}^{\mathrm{AF}}(v_{\mathrm{com}})
   = 0.0347\,v_{\mathrm{com}}^{2.08}\)) and consistent speed (\(\widehat{P}_{t,\mathrm{nd}}^{\mathrm{CS}}(v_{\mathrm{com}})
   = 0.0211\,v_{\mathrm{com}}^{2.49}\)) swimming were used to model the relationship between speed and power output. Power estimates during consistent speed swimming were lower than active fluking for the range of overlapping observed speeds. For an average swimming speed of 1.5 BL/s, the power estimates for active fluking were 33 percent higher. Average speeds during the active fluking were lower than those observed during consistent speed swimming for the three animals.

\begin{figure}[htbp]
    \centering
    \includegraphics[width=0.49\textwidth]{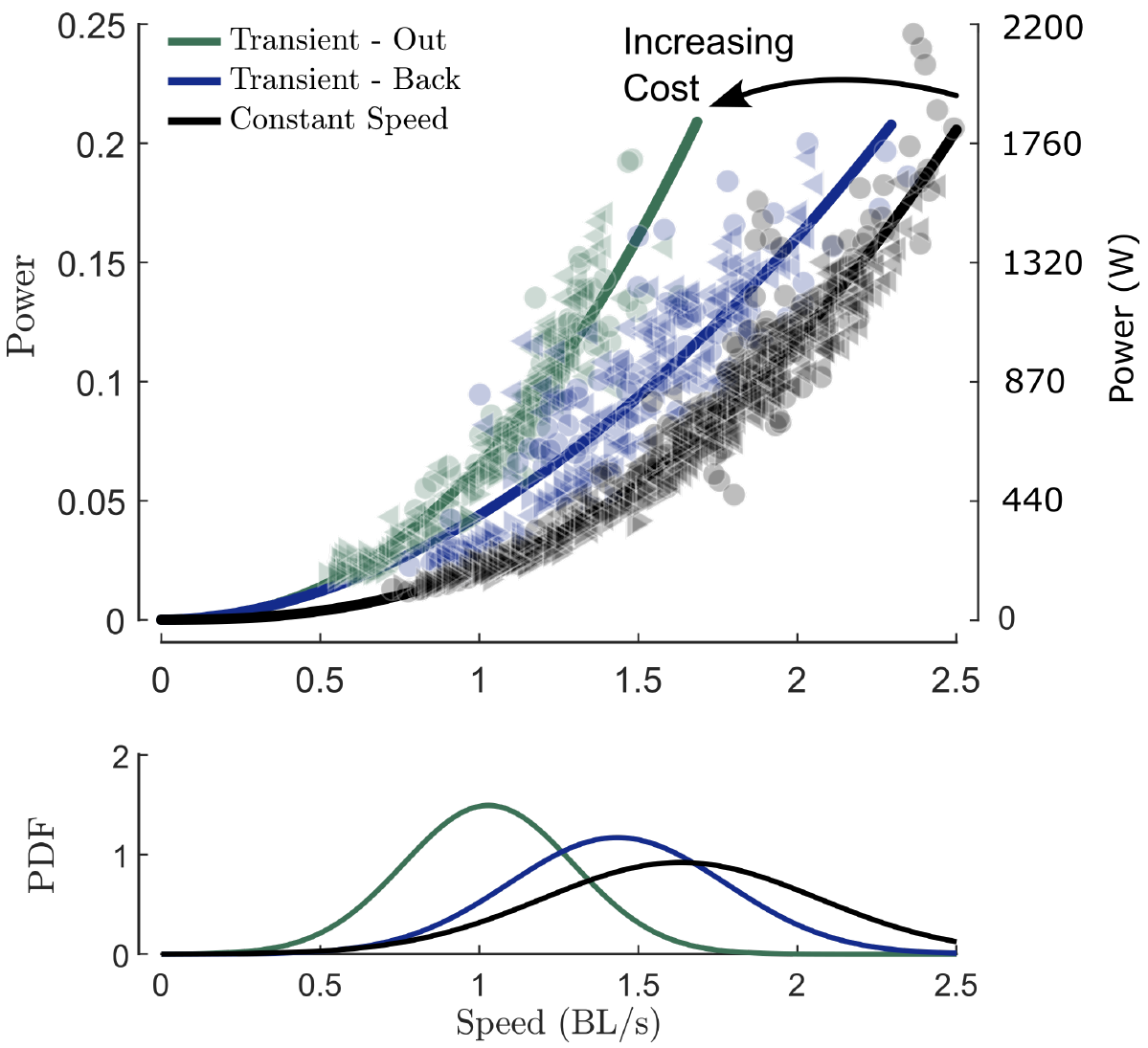}
    \caption{\textbf{Top:} Average power vs speed for periods of continuous and transient swimming. Average power for a given speed was largest during the first leg of the lap when the animal accelerated from station to a continuous swimming speed. \textbf{Bottom:} Speed distributions indicating that the animals' average swimming speed was fastest during the continuous swimming segments.}
    \label{fig:Trans VS Consistent Speed Power}
\end{figure}

The power output of the animals during the active fluking was the result of the propulsive force generated during both transient and consistent speed portions of each segment of the lap. Fig.~\ref{fig:Trans VS Consistent Speed Power} presents the average power output for the transient and consistent speed sections of each leg of the lap. Propulsive power was lowest during the consistent speed portions of each lap, and the estimated power during the initial transient portion of the lap was the largest. Power output during the second transient period after the turning event fell between the two periods. The average swimming speed observed during the consistent speed swimming (1.6 BL/s) was the fastest, whereas the slowest speed (1.0 BL/s) occurred during the initial transient phase.

\begin{figure}[htbp]
    \centering
    \includegraphics[width=0.49\textwidth]{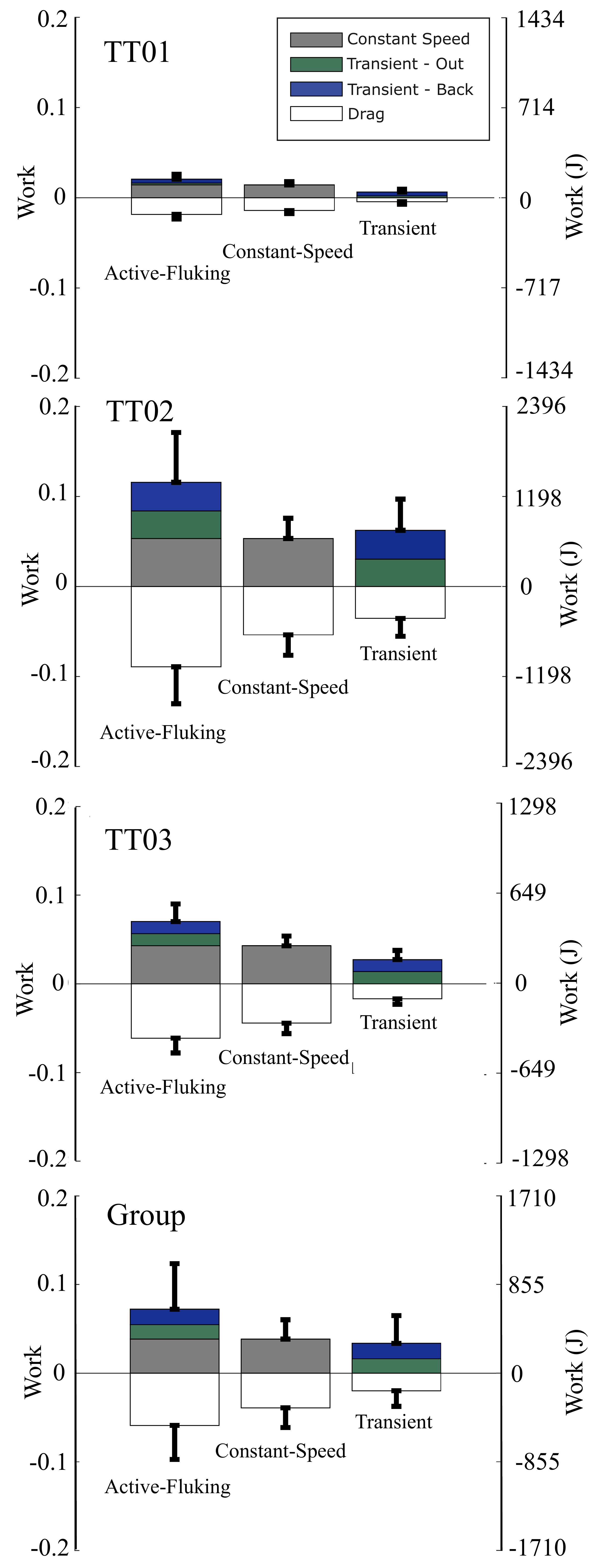}
    \caption{Non-dimensional work calculated from the thrust and drag power for three swimming phases: active fluking, transient swimming, and consistent speed swimming. The work during active fluking is the sum of the work done by the animal during the transient and consistent speed swimming.}
    \label{fig:work box plot by dolphin}
\end{figure}

The propulsive work during periods of active-fluking, consistent speed, and transient swimming was calculated for the dolphins, Fig.~\ref{fig:work box plot by dolphin}. This figure also presents an estimate of the negative work generated drag acting on the animals. During the consistent speed swimming, the propulsive work needed to maintain the swimming speed was equal to the estimated work due to drag, but during transient periods of swimming, the animals generated work to overcome drag and to accelerate the body. The work performed by TT02 was the highest of the three animals, and the work during the transient portion of the laps was higher than the work done during the consistent speed swimming. Work done by TT01 was the smallest during all three periods. The slower swimming speeds and lower acceleration in TT01's results account for much smaller estimates of work during the laps. The work done by TT03 fell between the two other animals, but more work was done during the consistent speed swimming compared to the transient.

\section{Discussion}\label{sec:Discussion}
Trade-offs between performance and cost can be used to help explain movement patterns of biological systems. During daily life, reducing energetic cost may be a primary driver in the selection of movement strategies. For example, during walking, humans have been observed to select speeds and gait parameters (like step-width) that minimize energetic cost \citep{abe2021economical}, \citep{abram2019energy}. In other contexts, performance and energetics may drive path planning and movement strategies. In a race, runners are balancing performance, completing a specified distance as quickly as possible, while balancing energetic cost within a range so that a race like a marathon can be finished. In this context, selecting a path to minimize the distance of the race (running on the inside of the track) or modulating power output to balance aerobic and anaerobic levels of exercise may drive decision-making. In this work, we were able to explore these trade-offs in bottlenose dolphins during a lap swimming task.

For the laps, both the time to complete the lap and the estimated energetic cost during the movement were used to investigate performance. Our results demonstrate that these dolphins exhibit significant individual differences in performance during the same task: swimming around a fixed marker in the lagoon. There were differences in lap time, path selection, and energetic cost of the movement patterns during both straight-line and cornering portions of the lap.  TT02 was the fastest dolphin during the task (\(23.1 \pm 2.6\) s) and had the fastest average peak speed (approximately 5.2 m/s). In contrast, TT01 took 43 percent more time to complete the lap (\(35.7 \pm 5.1\) s) with peak speeds that were 57 percent slower (approximately 2.9 m/s). Path selection also varied between the animals. TT02's average lap was \(78 \pm 8\) m long with a 1.9 m radius at the cornering event, TT03's average lap was \(78 \pm 4\) m long with a 1.0 m radius at the cornering event, and TT01's average lap was \(82 \pm 6\) m long with a 1.3 m radius at the cornering event, Fig.~\ref{fig:dr_xyt}. All of the animals swam fastest during the straight-line portions of the lap, and all slowed down during the cornering event. However, individual differences in speed and path selection significantly impacted estimated energetic cost.  

The speed and acceleration in the direction of motion are key kinematic input parameters for the model used to estimate the propulsive force generated by the animals during swimming. Energetic cost will be lower during periods of consistent speed swimming and will increase when the animals accelerate to faster swimming speeds. TT02 had the fastest measured swimming speeds and the largest periods of acceleration in the direction of motion, resulting in the highest estimates of propulsive power over most of the lap. Propulsive power for both TT01 and TT02 peaked just after the cornering event (approximately 60 percent of the lap, Fig.~\ref{fig:avg_trails}). All of the animals reduced their speed during the cornering, but TT02 had the largest relative change in speed, slowing down from a peak average speed of 4.6 m/s during the outgoing portion of the lap to 3.4 m/s at the cornering event. TT02 then accelerated to the fastest average peak swimming speed (5.2 m/s) of the three animals at approximately 60 percent of the lap. This combination of fast swimming speed and acceleration resulted in large propulsive power. In contrast, TT01 had the slowest swimming speeds (before and after the cornering event) and the smallest acceleration at the beginning of the return leg of the lap, both of which resulted in the lowest propulsive power estimates throughout the lap.

During lap swimming, the animals had periods of acceleration and constant speed swimming for each leg of the lap. The average thrust power output and cost of transport during active swimming (transient and consistent speed swimming) were higher than the periods of consistent speed swimming alone, Fig.~\ref{fig:AF VS CS COT}. This trend was consistent across the range of observed swimming speeds (0.6 - 2.5 BL/s). Even though the periods of transient motion only made up 20 percent of the active swimming trials, the additional power needed to accelerate the animal increased the cost of transport by an average of 32 percent over the range of average observed swimming speeds. For example, at an average speed of 1.5~BL/s, the dolphins had to increase the average thrust power output by 33 percent when the transient movement was included in the analysis. The additional propulsive thrust generated during the transient movement had a corresponding increase in the cost of transport, 38 percent at 1.5~BL/s. 

Managing acceleration during transient periods can reduce cost during swimming. The highest transient power output occurs at the beginning of the lap when the dolphins are accelerating from rest to a consistent swimming speed, Fig.~\ref{fig:Trans VS Consistent Speed Power}. At 1.5~BL/s, the power output during the transient period at the beginning of the lap was 94 percent larger than during the middle of the lap when the animal had reached a consistent speed of swimming. Power output during transient motion following the cornering event was also higher than during consistent speed swimming (48 percent higher at 1.5~BL/s), but lower than the initial period of transient motion (53 percent lower at 1.5~BL/s). These lower transient costs after the turning event are related to the maneuverability of the dolphin. 

The different movement strategies used by the dolphins before, during, and after the cornering event impacted overall cost. Maintaining speed through the cornering event reduces the transient cost at the start of the return leg of the lap, but the magnitude of the acceleration after the corner varied between the animals.  TT01 swam at the slowest speed and had the second largest radius of curvature at the cornering event, Fig.~\ref{fig:R_circle}. This slow swimming speed around a wider corner enabled TT01 to maintain a consistent swimming speed during the cornering event, Fig.~\ref{fig:avg_trails}. In contrast, TT02 had the fastest swimming speed and the widest turn at the cornering event (1.9 m radius), but lost the most speed swimming around the corner (around 2 m/s). TT03 had a slower average swimming speed than TT02 (2.8 m/s vs 3.3 m/s), but had the tightest turning radius (1.0 m) and a smaller change in speed (around 1 m/s) after the cornering event. 

These results indicate that individual animals may be weighing performance objectives differently. TT01 used a movement strategy that prioritized energetic cost (148~$J$ per lap) over lap performance (35.7 s lap duration). Because propulsive power output scales as speed cubed the slower swimming speeds significantly reduced the estimated work during the laps. The work done by TT01 was 102 percent smaller than TT03 (456~$J$), and 161 percent smaller than TT02 (1388~$J$). In contrast, TT02 prioritized performance (23.1~$s$ lap durations) over energetic cost. TT02 was the fastest animal (reaching average peak speeds of 5.2 m/s), and did the most work during both the transient (748 $J$) and consistent speed (640 $J$) portions of the laps. Unlike the other two dolphins, transient costs for TT02 were larger than the costs during the consistent speed swimming (748 $J$ vs 640 $J$). These costs reflect a performance-focused strategy where TT02 prioritizes fast peak speeds during the outgoing and return portions of the lap. The path selected by TT02 around the corner was the most variable and had the largest radius (1.9 m), but the faster peak speeds reached before and after the cornering event resulted in the larger transient costs.  

TT03 also used a movement strategy that prioritized performance, but demonstrated more maneuverability than TT02. During the consistent speed swimming the animals reached comparable peak swimming speeds (TT02 5.2 m/s and TT03 4.2 m/s) and had similar average cost during the consistent speed portions of the laps (TT02: 0.053~$norm.~work$ and TT03: 0.043~$norm.~work$), but the transient costs for TT03 (0.027~$norm.~work$) were 79 percent lower than for TT02 (0.062~$norm.~work$).  The path used for TT03 had a relatively tight cornering radius compared to TT02 (1 m vs 1.9 m) (Fig.~\ref{fig:dr_xyt}), and TT03 had the highest peak normal acceleration of the animals (6.0~$m/s^2$) as the animal maintained speed through the corner. This maneuverability around the corner resulted in lower transient costs on the return leg (Fig.~\ref{fig:avg_laps}). This combination of tight cornering with high acceleration enabled the rapid recovery to constant‐speed swimming to maintain performance (fast swimming times around a tight lap) without the increased energetic cost generated by transient movement.

In contrast with the lap swimming, observed movement strategies by the same animals during self-selected swimming prioritize energetic cost over performance~\citep{zhang2023dynamics},\citep{gabaldon2022tag}.  The lap swimming task performed by the three animals was a series of short-duration sprints (at the minute scale) compared to the longer periods of selected swimming (at the hour scale). A continuous fluking gait was used to reach mean peak swimming speeds that ranged from 2.9 to 5.2 m/s for the three animals during the laps. During extended periods of swimming (greater than an hour), the animals selected slower average speeds (around 1 m/s) and used gait patterns (fluke and glide) that minimized energetic cost \citep{gabaldon2022tag}. The fluke and glide gait consists of bursts of fluking to accelerate the animal, followed by periods of deceleration during the glide. The transient costs during the fluking to accelerate the body at these slower mean swimming speeds (approximately 1 m/s) are offset by the reduction in movement costs during the glide~\citep{zhang2023dynamics}. A fluke and glide gait was not observed during the lap swimming, indicating that there is a swimming speed where the animals transition from fluke and glide to continuous fluking that could be explained by the higher costs of transient movement.
\section{Conclusion}\label{sec:Conclusion}
Movement strategies selected by dolphins during swimming are likely to balance performance and energetic cost tradeoffs, and insight into how the animals reduce cost while maintaining performance can benefit the design and control of engineered systems. This work used a combination of kinematic measurements and physics-based models to investigate swimming biomechanics and energetics during a lap swimming task. Measured kinematics (speed and heading) were used to estimate the swimming path of the animal, and the models were used to estimate energetic cost (work and power) during lap swimming trials. Performance during the task was evaluated using a combination of the tag‐derived kinematics (speed, acceleration, orientation, and depth), the reconstructed swimming paths, and the molded swimming energetics. TT01 had the lowest energetic cost during the lap swimming but was also the slowest of the three animals, indicating that this dolphin selected a strategy that prioritized cost over performance. TT02 was the fastest lap swimmer but had the highest energetic cost. In contrast, TT03 completed laps at a slightly slower speed than TT02 but at half the estimated per-lap cost. This difference in efficiency was related to a reduction of transient costs during the lap. Specifically, TT03 selected a cornering strategy that reduced transient costs at the cornering event by selecting a cornering radius that enabled TT03 to better maintain its speed through the turn than TT02. TT03 is smaller and lighter than TT02, indicating that maneuverability during cornering is important for reducing transient costs. These results offer insights into the maneuverability and movement strategies dolphins employ to optimize the balance between performance and energetic costs. Additionally, trajectory selection and cornering behaviors observed with the dolphins can be used to develop control strategies to enhance path-planning techniques for robotic systems.

\section*{Acknowledgments}
This work was supported by a Contribution Agreement with the Department of Fisheries and Oceans Canada (DFO), and the National Science Foundation under Grant No. 2238432. The data used for the analysis will be provided upon request by contacting the corresponding author. This animal study was approved by the Institutional Animal Welfare and Use Committee at the University of Michigan.
The author would like to express sincere gratitude to Dolphin Quest Oahu for providing invaluable support to this research, including access to outstanding facilities and equipment. Special thanks are also extended to all the dedicated dolphin trainers and the cooperative dolphins who made this work possible. 
The authors declare no conflict of interest.

\clearpage
\bibliography{ref}
\end{document}